\errorstopmode
\input amssym.def
\input amssym.tex


\magnification=\magstephalf
\hsize=14.0 true cm
\vsize=19 true cm
\hoffset=1.0 true cm
\voffset=2.0 true cm

\abovedisplayskip=12pt plus 3pt minus 3pt
\belowdisplayskip=12pt plus 3pt minus 3pt
\parindent=1.0em


\font\sixrm=cmr6
\font\eightrm=cmr8
\font\ninerm=cmr9

\font\sixi=cmmi6
\font\eighti=cmmi8
\font\ninei=cmmi9

\font\sixsy=cmsy6
\font\eightsy=cmsy8
\font\ninesy=cmsy9

\font\sixbf=cmbx6
\font\eightbf=cmbx8
\font\ninebf=cmbx9

\font\eightit=cmti8
\font\nineit=cmti9

\font\eightsl=cmsl8
\font\ninesl=cmsl9

\font\sixss=cmss8 at 8 true pt
\font\sevenss=cmss9 at 9 true pt
\font\eightss=cmss8
\font\niness=cmss9
\font\tenss=cmss10

\font\sixmib=cmmib6
\font\sevenmib=cmmib7
\font\eightmib=cmmib8
\font\ninemib=cmmib9
\font\tenmib=cmmib10

 at 12 true pt
 at 12 true pt
\font\bigrm=cmr10 at 12 true pt
 at 12 true pt
 at 12 true pt

 at 16 true pt
 at 16 true pt
\font\Bigrm=cmr12 at 16 true pt
 at 16 true pt
 at 16 true pt

\catcode`@=11
\newfam\ssfam
\newfam\mibfam

\def\tenpoint{\def\rm{\fam0\tenrm}%
    \textfont0=\tenrm \scriptfont0=\sevenrm \scriptscriptfont0=\fiverm
    \textfont1=\teni  \scriptfont1=\seveni  \scriptscriptfont1=\fivei
    \textfont2=\tensy \scriptfont2=\sevensy \scriptscriptfont2=\fivesy
    \textfont3=\tenex \scriptfont3=\tenex   \scriptscriptfont3=\tenex
    \textfont\itfam=\tenit                  \def\it{\fam\itfam\tenit}%
    \textfont\slfam=\tensl                  \def\sl{\fam\slfam\tensl}%
    \textfont\bffam=\tenbf \scriptfont\bffam=\sevenbf
                           \scriptscriptfont\bffam=\fivebf
                           \def\bf{\fam\bffam\tenbf}%
    \textfont\ssfam=\tenss \scriptfont\ssfam=\sevenss
                           \scriptscriptfont\ssfam=\sevenss
                           \def\ss{\fam\ssfam\tenss}%
    \textfont\mibfam=\tenmib \scriptfont\mibfam=\sevenmib
                             \scriptscriptfont\mibfam=\sevenmib
                             \def\mib{\fam\mibfam\tenmib}%
    \normalbaselineskip=13pt
    \setbox\strutbox=\hbox{\vrule height8.5pt depth3.5pt width0pt}%
    \let\big=\tenbig
    \normalbaselines\rm}

\def\ninepoint{\def\rm{\fam0\ninerm}%
    \textfont0=\ninerm      \scriptfont0=\sixrm
                            \scriptscriptfont0=\fiverm
    \textfont1=\ninei       \scriptfont1=\sixi
                            \scriptscriptfont1=\fivei
    \textfont2=\ninesy      \scriptfont2=\sixsy
                            \scriptscriptfont2=\fivesy
    \textfont3=\tenex       \scriptfont3=\tenex
                            \scriptscriptfont3=\tenex
    \textfont\itfam=\nineit \def\it{\fam\itfam\nineit}%
    \textfont\slfam=\ninesl \def\sl{\fam\slfam\ninesl}%
    \textfont\bffam=\ninebf \scriptfont\bffam=\sixbf
                            \scriptscriptfont\bffam=\fivebf
                            \def\bf{\fam\bffam\ninebf}%
    \textfont\ssfam=\niness \scriptfont\ssfam=\sixss
                            \scriptscriptfont\ssfam=\sixss
                            \def\ss{\fam\ssfam\niness}%
    \textfont\mibfam=\ninemib \scriptfont\mibfam=\sixmib
                            \scriptscriptfont\mibfam=\sixmib
                            \def\mib{\fam\mibfam\ninemib}%
    \normalbaselineskip=12pt
    \setbox\strutbox=\hbox{\vrule height8.0pt depth3.0pt width0pt}%
    \let\big=\ninebig
    \normalbaselines\rm}

\def\eightpoint{\def\rm{\fam0\eightrm}%
    \textfont0=\eightrm      \scriptfont0=\sixrm
                             \scriptscriptfont0=\fiverm
    \textfont1=\eighti       \scriptfont1=\sixi
                             \scriptscriptfont1=\fivei
    \textfont2=\eightsy      \scriptfont2=\sixsy
                             \scriptscriptfont2=\fivesy
    \textfont3=\tenex        \scriptfont3=\tenex
                             \scriptscriptfont3=\tenex
    \textfont\itfam=\eightit \def\it{\fam\itfam\eightit}%
    \textfont\slfam=\eightsl \def\sl{\fam\slfam\eightsl}%
    \textfont\bffam=\eightbf \scriptfont\bffam=\sixbf
                             \scriptscriptfont\bffam=\fivebf
                             \def\bf{\fam\bffam\eightbf}%
    \textfont\ssfam=\eightss \scriptfont\ssfam=\sixss
                             \scriptscriptfont\ssfam=\sixss
                             \def\ss{\fam\ssfam\eightss}%
    \textfont\mibfam=\eightmib \scriptfont\mibfam=\sixmib
                             \scriptscriptfont\mibfam=\sixmib
                             \def\mib{\fam\mibfam\eightmib}%
    \normalbaselineskip=10pt
    \setbox\strutbox=\hbox{\vrule height7.0pt depth2.0pt width0pt}%
    \let\big=\eightbig
    \normalbaselines\rm}

\def\tenbig#1{{\hbox{$\left#1\vbox to8.5pt{}\right.\n@space$}}}
\def\ninebig#1{{\hbox{$\textfont0=\tenrm\textfont2=\tensy
                       \left#1\vbox to7.25pt{}\right.\n@space$}}}
\def\eightbig#1{{\hbox{$\textfont0=\ninerm\textfont2=\ninesy
                       \left#1\vbox to6.5pt{}\right.\n@space$}}}

\font\sectionfont=cmbx10
\font\subsectionfont=cmti10

\def\figurecaptionfont{\ninepoint}
\def\tablecaptionfont{\ninepoint}
\def\footnotefont{\eightpoint}


\newcount\equationno
\newcount\bibitemno
\newcount\figureno
\newcount\tableno

\equationno=0
\bibitemno=0
\figureno=0
\tableno=0


\footline={\ifnum\pageno=0{\hfil}\else
{\hss\rm\the\pageno\hss}\fi}


\def\section #1. #2 \par
{\vskip0pt plus .10\vsize\penalty-100 \vskip0pt plus-.10\vsize
\vskip 1.6 true cm plus 0.2 true cm minus 0.2 true cm
\global\def\equationlabel{#1}
\global\equationno=0
\leftline{\sectionfont #1. #2}\par
\immediate\write\terminal{Section #1. #2}
\vskip 0.7 true cm plus 0.1 true cm minus 0.1 true cm
\noindent}


\def\subsection #1 \par
{\vskip0pt plus 1.0 true cm\penalty-50 \vskip0pt plus-1.0 true cm
\vskip2.5ex plus 0.1ex minus 0.1ex
\leftline{\subsectionfont #1}\par
\immediate\write\terminal{Subsection #1}
\vskip1.0ex plus 0.1ex minus 0.1ex
\noindent}


\def\appendix #1. #2 \par
{\vskip0pt plus .10\vsize\penalty-100 \vskip0pt plus-.10\vsize
\vskip 1.6 true cm plus 0.2 true cm minus 0.2 true cm
\global\def\equationlabel{\hbox{\rm#1}}
\global\equationno=0
\leftline{\sectionfont Appendix #1. #2}\par
\immediate\write\terminal{Appendix #1. #2}
\vskip 0.7 true cm plus 0.1 true cm minus 0.1 true cm
\noindent}



\def\equation#1{$$\displaylines{\qquad #1}$$}
\def\enum{\global\advance\equationno by 1
\hfill\llap{{\rm(\equationlabel.\the\equationno)}}}
\def\noenum{\hfill}

\def\nexteq#1{\cr\noalign{\vskip#1}\qquad}


\def\ifundefined#1{\expandafter\ifx\csname#1\endcsname\relax}

\def\ref#1{\ifundefined{#1}?\immediate\write\terminal{unknown reference
on page \the\pageno}\else\csname#1\endcsname\fi}

\newwrite\terminal
\newwrite\bibitemlist

\def\bibitem#1#2\par{\global\advance\bibitemno by 1
\immediate\write\bibitemlist{\string\def
\expandafter\string\csname#1\endcsname
{\the\bibitemno}}
\item{[\the\bibitemno]}#2\par}

\def\beginbibliography{
\vskip0pt plus .15\vsize\penalty-100 \vskip0pt plus-.15\vsize
\vskip 1.2 true cm plus 0.2 true cm minus 0.2 true cm
\leftline{\sectionfont References}\par
\immediate\write\terminal{References}
\immediate\openout\bibitemlist=biblist
\frenchspacing\parindent=1.8em
\vskip 0.5 true cm plus 0.1 true cm minus 0.1 true cm}

\def\endbibliography{
\immediate\closeout\bibitemlist
\nonfrenchspacing\parindent=1.0em}


\def\figurecaption#1{\global\advance\figureno by 1
\narrower\figurecaptionfont Fig.~\the\figureno. #1}

\def\tablecaption#1{\global\advance\tableno by 1
\centerline{\tablecaptionfont Table~\the\tableno. #1}}

\def\thicktablerule{\hrule height0.8pt}
\def\thintablerule{\hrule height0.4pt}

\tenpoint

\immediate\openin\bibitemlist=biblist
\ifeof\bibitemlist\immediate\closein\bibitemlist
\else\immediate\closein\bibitemlist
\input biblist \fi


\def\thismonth{\ifcase\month\or
January\or February\or March\or April\or May\or June\or
July\or August\or September\or October\or November\or December\fi}

\input epsf
\epsfclipon



\def\rmd{{\rm d}}

\def\rme{{\rm e}}
\def\rmO{{\rm O}}



\def\proof{\noindent{\sl Proof:}\kern0.6em}

\def\frac#1#2{\hbox{$#1\over#2$}}
\def\dual{\mathstrut^*\kern-0.1em}
\def\mod{\;\hbox{\rm mod}\;}

\def\lvec#1{\setbox0=\hbox{$#1$}
    \setbox1=\hbox{$\scriptstyle\leftarrow$}
    #1\kern-\wd0\smash{
    \raise\ht0\hbox{$\raise1pt\hbox{$\scriptstyle\leftarrow$}$}}
    \kern-\wd1\kern\wd0}
\def\rvec#1{\setbox0=\hbox{$#1$}
    \setbox1=\hbox{$\scriptstyle\rightarrow$}
    #1\kern-\wd0\smash{
    \raise\ht0\hbox{$\raise1pt\hbox{$\scriptstyle\rightarrow$}$}}
    \kern-\wd1\kern\wd0}
\def\cvec#1{\kern-0.5pt\vec{\kern0.5pt #1}}

\def\slash#1{\setbox2=\hbox{$\displaystyle#1$}%
             \setbox3=\hbox{$\displaystyle/$}%
             #1\kern-0.8\wd2/\kern-1.0\wd3\kern0.8\wd2\kern0.5pt}

\def\wick#1{\setbox2=\hbox{$\displaystyle#1$}
    \setbox3=\null\ht3=3.0pt\dp3=0.0pt\wd3=20.0pt
    #1\kern-\wd2\kern3.0pt\raise11.0pt\vbox{\hrule height0.3pt
    \hbox{\vrule width0.3pt\box3\vrule width0.3pt}}\kern-24.0pt\kern\wd2}

\def\longwick#1{\setbox2=\hbox{$\displaystyle#1$}
    \setbox3=\null\ht3=3.0pt\dp3=0.0pt\wd3=27.0pt
    #1\kern-\wd2\kern3.0pt\raise11.0pt\vbox{\hrule height0.3pt
    \hbox{\vrule width0.3pt\box3\vrule width0.3pt}}\kern-31.0pt\kern\wd2}

\def\verylongwick#1{\setbox2=\hbox{$\displaystyle#1$}
    \setbox3=\null\ht3=3.0pt\dp3=0.0pt\wd3=43.0pt
    #1\kern-\wd2\kern3.0pt\raise11.0pt\vbox{\hrule height0.3pt
    \hbox{\vrule width0.3pt\box3\vrule width0.3pt}}\kern-47.0pt\kern\wd2}


\def\nabstar#1{{\nabla\kern0.5pt\smash{\raise 4.5pt\hbox{$\ast$}}
               \kern-5.5pt_{#1}}}
\def\drv#1{{\partial_{#1}}}
\def\drvstar#1{{\partial\kern0.5pt\smash{\raise 4.5pt\hbox{$\ast$}}
               \kern-6.0pt_{#1}}}
\def\sdrvstar#1{{\partial\kern0.4pt\smash{\raise 3.6pt\hbox{$\ast$}}
                \kern-4.8pt_{#1}}}

\def\ldrvstar#1{{\lvec{\,\partial}\kern-0.5pt\smash{\raise 4.5pt\hbox{$\ast$}}
               \kern-5.0pt_{#1}}}


\def\MSbar{\overline{\rm MS\kern-0.5pt}\kern0.5pt}



\def\chibar{\overline{\chi}}

\def\ren#1{#1_{\hbox{\sixrm R}}}


\def\diracstar#1#2{
    \setbox0=\hbox{$\gamma$}\setbox1=\hbox{$\gamma_{#1}$}
    \gamma_{#1}\kern-\wd1\kern\wd0
    \smash{\raise4.5pt\hbox{$\scriptstyle#2$}}}


\def\SUthree{{\rm SU(3)}}
\def\SUn{{\rm SU}(N)}

\def\tr{{\rm tr}}


\def\SG{S_{\rm G}}
\def\Sw{S_{\rm w}}
\def\Sgf{S_{\rm gf}}
\def\lp{{\cal C}}
\def\lps#1{{\cal S}_{#1}}


\def\Dw{D_{\rm w}}
\def\Dwdag{{\Dw}\kern-4pt^{\dagger}\kern1pt}
\def\Dm{D}
\def\Dmdag{\Dm^{\dagger}\kern-1pt}

\def\Dmsdag{D_s\kern-2pt\vphantom{D}^{\dagger}\kern-1pt}


\def\csw{c_{\rm sw}}
\def\cG{c_{\hbox{\sixrm G}}}

\def\ca{c_A}


\def\Nf{N_{\rm f}}

\def\gbar{\bar{g}}
\def\gbarinf{\gbar_{\infty}}
\def\kfact{k}
\def\kfactinf{\kfact_{\infty}}
\def\tauint{\tau_{\rm int}}
\def\openSF{open-SF }
\def\sump#1#2{\sum_{#1}^{#2}\kern-2pt{\vphantom{\sum}}'}
\def\Ebar{\kern1.5pt\overline{\kern-1.5ptE\kern-0.5pt}\kern0.5pt}
\def\obs{{\cal O}}

%
\rightline{CERN-PH-TH/2014-065}
\vskip1.2cm
\centerline{\Bigrm Step scaling and the Yang--Mills gradient flow}

\vskip 0.6 true cm
\centerline{\bigrm Martin L\"uscher}
\vskip1.5ex
\centerline{{\it CERN, Physics Department, 1211 Geneva 23, Switzerland}}
\vskip 0.8 true cm
\thintablerule
\vskip 2.0ex
\ninepoint
\leftline{\bf Abstract}
\vskip 1.0ex\noindent
The use of the Yang--Mills gradient
flow in step-scaling studies of lattice QCD
is expected to lead to results of unprecedented precision.
Step scaling is usually based on the Schr\"odinger functional,
where time ranges over an interval $[0,T]$ and all fields
satisfy Dirichlet boundary conditions at time $0$ and $T$.
In these calculations,
potentially important sources of systematic errors
are boundary lattice effects and the infamous topology-freezing problem.
The latter is here shown to be absent if Neumann instead
of Dirichlet
boundary conditions are imposed on the gauge field at time $0$.
Moreover, the expectation values of gauge-invariant
local fields at positive flow time
(and of other well localized observables) that reside in the center
of the space-time volume
are found to be largely insensitive to the
boundary lattice effects.

\vskip 2.0ex
\thintablerule

\tenpoint


\section 1. Introduction

In numerical lattice field theory, step scaling refers to
a finite-size scaling technique that allows the physics
at high energies to be related to the characteristic
low energy scales of the theory [\ref{StepScaling}].
While no uncontrolled approximations need to be made
in these calculations, the precision of the results that
can be reached in practice depends on many details.
In particular, the boundary conditions in finite volume
and the observables should be such that the systematic and
statistical errors are minimized.

Using the Yang--Mills gradient flow
[\ref{WilsonFlow}--\ref{ChFlow}],
many observables can be constructed that
are expected to be well suited for step-scaling studies.
The expectation value of the square of
the gauge-field tensor at positive flow time, for example,
can serve as running coupling in finite volume
[\ref{NogradiEtAl}--\ref{JarnoMWT}].
Another motivation for considering such observables
derives from their small flow-time expansion
in local fields [\ref{RenFlow},\ref{EMflowI}--\ref{Mainz}].
Over a wide range of flow time,
the coefficients in these expansions may conceivably be computed
using an adapted step-scaling technique.
Representations of local fields would then be
obtained, where the field renormalization and Symanzik-improvement
are greatly simplified.

In lattice QCD, step scaling is usually based on
the Schr\"odinger functional [\ref{SF},\ref{SFquark}].
There are good reasons for this choice, among them
the absence of zero-modes in perturbation theory,
the renormalizability of the Schr\"odinger functional
and the fact that
the theory can be probed by varying the boundary values of the gauge field.
On the other hand,
the boundaries of the lattice in the time direction
are a source of additional lattice effects.
Moreover, near the continuum limit, simulations of the
Schr\"odinger functional tend to get trapped in one of the emergent
sectors of fixed topological charge of the gauge field.

Until recently [\ref{FixedQ}], topology-freezing was
not thought to be an issue in step-scaling studies.
In small space-time volumes,
where most of the step scaling is performed,
the non-trivial topological sectors are in fact
strongly suppressed.
The width of the (unbiased) charge distribution
however grows rapidly towards the non-perturbative large-volume regime.
A proper sampling of the charge sectors is then required,
or a way of bypassing the problem
(such as restricting the theory
to the trivial sector [\ref{FixedQ}])
must be found.
Resolving this technical issue is particularly
urgent when observables defined at positive gradient-flow time
are used, since these can be rather sensitive
to autocorrelations in simulation time [\ref{openQCD}].

In this paper, a modification of the Schr\"odinger functional
is considered, where open (Neumann) boundary conditions are
imposed on the gauge field at one of the space-time boundaries
and Dirichlet boundary conditions at the opposite boundary.
With this choice, the renormalizability of the theory is preserved,
the perturbation expansion in powers of the gauge
coupling remains regular and the topology-freezing problem is avoided,
because the topological charge can freely flow in and out of the
volume through the open boundary [\ref{openQCD}].
The volume-dependence of the expectation values of observables
localized in the center of the space-time volume is then
examined and shown to be small in the kinematical situations of interest.
In particular, such quantities
are practically unaffected by boundary lattice effects.
Both limitations of the
Schr\"odinger functional setup
can thus be overcome with the suggested change of
boundary conditions and if suitable local observables
(such as the ones obtained with the gradient flow) are used.

\section 2. QCD with \openSF boundary conditions

The proposed modification of the Schr\"odinger functional
[\ref{SF},\ref{SFquark}] is rather modest and only concerns
the gauge field. In particular, the boundary conditions
imposed on the quark fields are left unchanged and need
not be discussed here.

\subsection 2.1 Continuum theory

As in the case of the Schr\"odinger functional,
the theory is set up on a four-dimen\-sion\-al
box with Euclidean metric,
time-like extent $T$ and size $L$ in the space directions.
Periodic boundary conditions are chosen in space
so that the space-time manifold has
two boundaries, one at, say,
time $x_0=0$ and the other at $x_0=T$.

For simplicity, only homogeneous boundary conditions are
considered here, where the gauge potential $A_{\mu}(x)$
is required to satisfy
\equation{
  \left.F_{0k}(x)\right|_{x_0=0}=0,
  \qquad
  F_{\mu\nu}(x)=\partial_{\mu}A_{\nu}-\partial_{\nu}A_{\mu}
  +[A_{\mu},A_{\nu}],
  \enum
  \nexteq{2.5ex}
  \left.A_k(x)\right|_{x_0=T}=0,   \qquad k=1,2,3
  \enum
}
(the notational conventions are summarized in appendix A).
The only difference
with respect to the Schr\"odinger functional
is the choice of Neumann rather than Dirichlet
boundary conditions at time $0$.
Neumann boundary conditions are also referred to as open boundary
conditions and the term ``\openSF boundary conditions'' is,
in the following, reserved for the combination
(2.1),(2.2) of boundary conditions. For brevity, homogenous
Dirichlet boundary conditions at both time $0$ and time $T$
will be referred to as ``SF boundary conditions".

With \openSF boundary conditions, the gauge potential
remains unconstrained at $x_0=0$ and only its (gauge-covariant) time derivative
must vanish. As a consequence,
the topological charge of the gauge field is no longer quantized and
can freely flow in and out of
the space-time volume through the boundary [\ref{openQCD}].
The renormalizability of the theory, on the other hand, is preserved,
since there are no candidate boundary counterterms of
dimension $3$ with the required symmetry properties
[\ref{SF},\ref{SFquark},\ref{openQCD}].

\subsection 2.2 Lattice formulation

In the lattice theory, space-time is replaced by
a hypercubic lattice of points $x$ with integer Cartesian coordinates
$x_0,x_1,x_2,x_3$ in the range
\equation{
  0\leq x_0\leq T,
  \qquad
  0\leq x_k<L\quad(k=1,2,3).
  \enum
}
For notational convenience, the lattice sizes $T,L$
and all other dimensionful quantities
are given in units of the lattice spacing $a$.
Periodic boundary conditions in space imply that
any two points $x$ and $x+\hat{k}\mod L$
(where $\hat{\mu}$ denotes the unit vector in direction $\mu$)
are considered to be nearest neighbors.

As usual the gauge field is represented by a field of
matrices $U(x,\mu)\in\SUn$ on the links $(x,x+\hat{\mu})$
of the lattice. No field variables are assigned to the
links that ``stick out of the lattice'',
i.e.~$\mu=0,1,2,3$ if $0\leq x_0<T$ and $\mu=1,2,3$ if $x_0=T$.
Imposing \openSF boundary conditions amounts to setting
\equation{
  \left.U(x,k)\right|_{x_0=T}=1,\qquad k=1,2,3,
  \enum
}
while all other field variables are left unconstrained and are
integrated over in the QCD functional integral.
In particular,
the boundary condition (2.1) does not
need to be enforced on the lattice, since it emerges
dynamically when the continuum limit is taken
[\ref{openQCD}].

\topinsert
\vbox{
\vskip0ex

\epsfxsize=9.0cm\hskip1.0cm\epsfbox{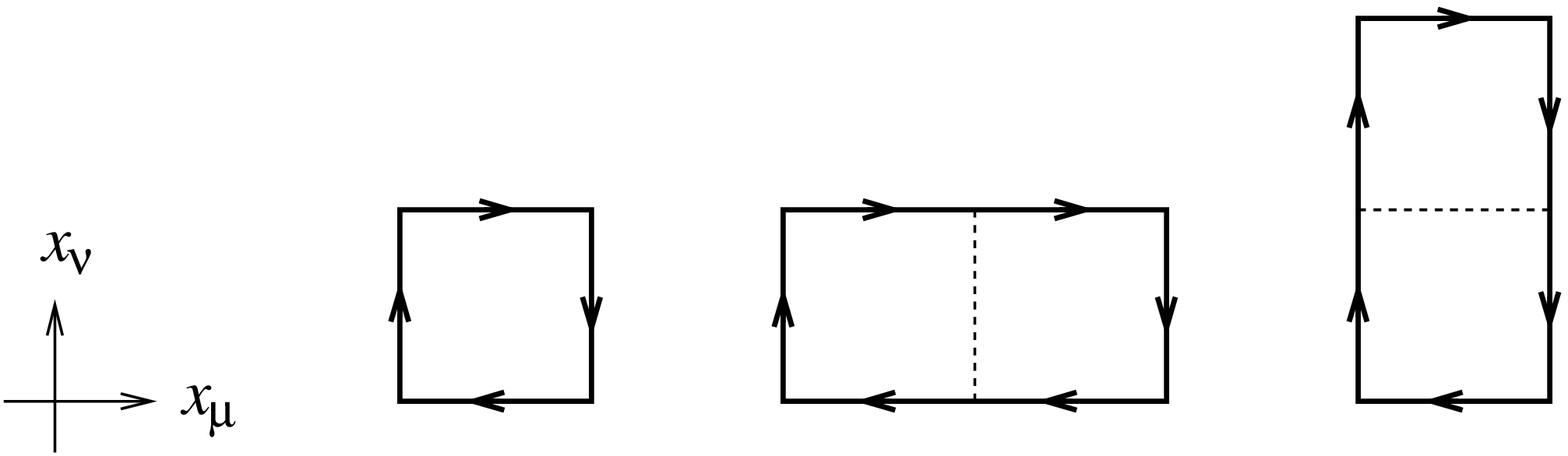}

\vskip1.5ex
\figurecaption{
Plaquette and planar double-plaquette loops
in a $(\mu,\nu)$--plane of the lattice.
Loops differing by their orientation
are considered to be different.}
}
\vskip0ex
\endinsert

\subsection 2.3 Lattice action

Let $\lps{0}$ and $\lps{1}$ be the sets of oriented
plaquette and double-plaquette loops on the lattice (see fig.~1).
Many popular lattice actions for the gauge field
are of the general form [\ref{Wilson}--\ref{OnShell}]
\equation{
  \SG={1\over g_0^2}\sum_{k=0}^1c_k\sum_{\lp\in\lps{k}}
  w_k(\lp)\,\tr\{1-U(\lp)\},
  \enum
}
where $U(\lp)$ denotes the ordered product of the link variables
around the loop $\lp$.
The weight factors $w_k(\lp)$ differ from
unity only near the boundaries of the
lattice and will be specified below.

In order to ensure the correct normalization
of the bare coupling $g_0$, the coefficients $c_k$ must be
such that
\equation{
  c_0+8c_1=1.
  \enum
}
Moreover, the constraint $c_0>0$ is imposed as otherwise
there may be fields with lowest action which are not locally pure-gauge
configurations [\ref{OnShell}].

\topinsert
\vbox{
\vskip0.0ex

\epsfxsize=7.5cm\hskip2.25cm\epsfbox{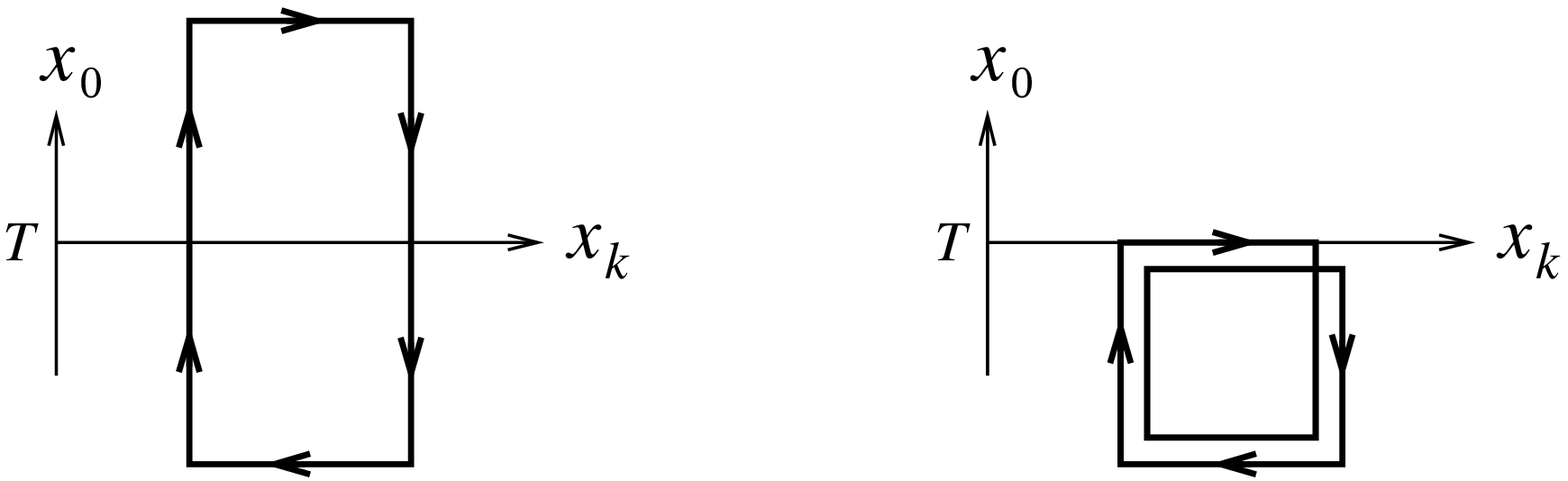}

\vskip2.0ex
\figurecaption{%
The time-like double-plaquette loops $\lp$ that cross the boundary
at time $T$ are included in the gauge action (2.5),
with weight $w_1(\lp)=1/2$
and $U(\lp)$ set to the product of the link variables around the
loop shown on the right.
}}
\vskip0.0ex
\endinsert

The sum in eq.~(2.5) runs over all loops $\lp$ that
are fully contained in the range $0\leq x_0\leq T$ of time.
In addition, the double-plaquette loops shown in fig.~2
are included in the sum.
The weights of the plaquette loops are then
\equation{
  w_0(\lp)=\cases{
  \frac{1}{2}\cG & if $\lp$ is a spatial loop at $x_0=0$,\cr
  \noalign{\vskip1.0ex}
  \cG'           & if $\lp$ has exactly one spatial link at $x_0=T$,\cr
  \noalign{\vskip1.0ex}
  1              & otherwise,\cr
  }
  \enum
}
while those of the double-plaquette loops are given by
\equation{
  w_1(\lp)=\cases{
  \frac{1}{2}\cG & if $\lp$ is a spatial loop at $x_0=0$,\cr
  \noalign{\vskip1.0ex}
  \frac{1}{2}    & if $\lp$ crosses the boundary at $x_0=T$ as in fig.~2,\cr
  \noalign{\vskip1.0ex}
  1              & otherwise,\cr
  }
  \enum
}
where $\cG$ and $\cG'$ are improvement coefficients that
should be tuned so as to cancel the O($a$) boundary lattice effects
in on-shell quantities [\ref{SF}].
At tree-level of perturbation theory, this is achieved by
setting $\cG=\cG'=1$%
\kern1pt\footnote{$\dagger$}{\footnotefont%
The form of the action near the boundary at $x_0=T$ differs from
both choice A and B proposed by Aoki, Frezzotti and Weisz [\ref{SFaction}].
Choice B is recommended for studies of
the Schr\"odinger functional with a non-trivial
background field, while A is the more convenient choice for perturbation
theory in the absence of a background field.
The action defined here combines the good properties
of A and B without having any obvious disadvantages.}.

\subsection 2.4 Stability at weak coupling

At small coupling $g_0$, the QCD functional integral is dominated
by the gauge-field configurations in the neighborhood of the
configurations with lowest action. The latter
can be shown to be gauge-equivalent to the classical
vacuum configuration, both in the continuum theory and in
the O($a$)-improved lattice theory [\ref{OnShell}].
Moreover, the Hessian of the action
at the classical vacuum configuration
has no zero modes other than the gauge modes
(see sect.~3).
With \openSF boundary conditions, the theory is thus
practically Gaussian in the weak-coupling domain and
is therefore expected to be entirely well-behaved
in this regime.

\section 3. Perturbation theory

The perturbation expansion of the theory with \openSF boundary
conditions does not run into any particular difficulties
and actually tends to be slightly simpler than in the case of
the Schr\"odinger functional.
In this section, the gauge-field propagator
is calculated in the O($a$)-improved lattice theory.
The result is obtained in a form in which the continuum limit
is transparent and that allows
the normalization of the gradient-flow coupling to be determined
in a few lines (see sect.~4).

\subsection 3.1 Fourier representation

In perturbation theory, the link
variables\kern1pt\footnote{$\dagger$}{\footnotefont%
For notational convenience, the same symbol is used for
the gauge potential as in the continuum theory.
Notice that the gauge potential is scaled by
the coupling $g_0$ in perturbation theory.}
\equation{
  U(x,\mu)=\exp\{g_0A_{\mu}(x)\}
  \enum
}
are parameterized through a lattice gauge potential
$A_{\mu}(x)$.
The latter can be decomposed into Fourier modes according to
\equation{
  A_0(x)={2i\over TL^3}\sum_p
  \sin(p_0x_0+\frac{1}{2}p_0)\rme^{i{\mib px}}\tilde{A}_0(p),
  \enum
  \nexteq{2.0ex}
  A_k(x)={2\over TL^3}\sum_p
  \cos(p_0x_0)\rme^{i{\mib px}+{i\over2}p_k}\tilde{A}_k(p),
  \enum
}
where the momentum components run over the values
\equation{
  p_0=(n_0+\frac{1}{2}){\pi\over T}, \qquad n_0=0,1,\ldots,T-1,
  \enum
  \nexteq{2.5ex}
  p_k=n_k{2\pi\over L},\qquad n_k=0,1,\ldots,L-1,\qquad k=1,2,3.
  \enum
}
Note that $p=0$ is not contained in the momentum spectrum
and that the Fourier representation (3.3) respects the
boundary condition (2.4) since $\cos(p_0T)=0$ for all momenta.
In view of the reality of the gauge potential in
position space,
the Fourier components must satisfy
\equation{
  \tilde{A}^a_0(p)^{\ast}=-\tilde{A}^a_0(p_0,-{\mib p}),
  \qquad
  \tilde{A}^a_k(p)^{\ast}=\tilde{A}^a_k(p_0,-{\mib p}),
  \enum
}
but are otherwise unconstrained.

Some algebra now shows that the gauge action (2.5)
[with $\cG,\cG'=1+\rmO(g_0^2)$] is given by
\equation{
  \SG={1\over 2TL^3}\sum_p
  \sum_{\mu,\nu}\bigl(1-c_1(\hat{p}_{\mu}^2+\hat{p}_{\nu}^2)\bigr)
  \bigl|\hat{p}_{\mu}\tilde{A}^a_{\nu}(p)-
        \hat{p}_{\nu}\tilde{A}^a_{\mu}(p)\bigr|^2+\rmO(g_0)
  \enum
}
to leading order in the gauge coupling, where
\equation{
  \hat{p}_{\mu}=2\sin(\frac{1}{2}p_{\mu}).
  \enum
}
The Fourier transformation thus diagonalizes the quadratic part of the
action. One might consider this to be hardly worth pointing out,
but whether the property holds or not actually depends on the exact form of
the action near the boundaries of the lattice [\ref{SFaction}].

\subsection 3.2 Gauge fixing

With the chosen boundary conditions,
it suffices to fix the subgroup of gauge transformations
$\Lambda(x)\in\SUn$ satisfying
\equation{
  \left.\Lambda(x)\right|_{x_0=T}=1.
  \enum
}
The global gauge transformations then remain a symmetry of the
gauge-fixed theory.
A possible choice of the gauge-fixing term is
\equation{
  \Sgf=\frac{1}{2}\lambda_0(\eth A,\eth A),
  \enum
  \nexteq{2.5ex}
  (\eth A)(x)=\sum_{\mu}(1+c_1\drvstar{\mu}\drv{\mu})\drvstar{\mu}A_{\mu}(x),
  \enum
}
where $\drv{\mu}$ and $\drvstar{\mu}$ denote the standard forward
and backward difference operators, $\lambda_0>0$ is the gauge-fixing
parameter and the terms proportional to $c_1$ are included in
order to simplify the expression for the gauge-field propagator
(see appendix B).
The scalar product in eq.~(3.10),
\equation{
  (\omega,\nu)=\sump{x_0=0}{T}\sum_{\mib x}
  \omega^a(x)\nu^a(x),
  \enum
}
is the one in the space of infinitesimal gauge transformations
(the primed summation symbol indicates that the terminal summands are
counted with weight $1/2$).

Near the boundaries of the lattice,
the derivatives in eq.~(3.11) require some further specification,
since the gauge potential $A_{\mu}(x)$ is only defined on
the links $(x,x+\hat{\mu})$ in the range $[0,T]$ of time.
A simple prescription for the boundary terms amounts
to extending the potential to all points $x$
with integer coordinates
through its Fourier representation (3.2),(3.3) so that
\equation{
  (\eth A)(x)=
  {2i\over TL^3}\sum_p\sum_{\mu}
  \cos(p_0x_0)\rme^{i{\mib px}}(\hat{p}_{\mu}-c_1\hat{p}_{\mu}^3)
  \tilde{A}_{\mu}(p).
  \enum
}
For the gauge-fixed action the formula
\equation{
  \SG+\Sgf={1\over TL^3}\sum_p\sum_{\mu,\nu}
  \tilde{A}^a_{\mu}(p)^{\ast}\Delta_{\mu\nu}(p)
  \tilde{A}^a_{\nu}(p)+\rmO(g_0),
  \enum
  \nexteq{2.5ex}
  \Delta_{\mu\nu}(p)=\delta_{\mu\nu}\hat{p}^2-\hat{p}_{\mu}\hat{p}_{\nu}
  -c_1\left\{\delta_{\mu\nu}
  \left[\hat{p}^4+\frac{1}{2}\hat{p}^2
  \left(\hat{p}_{\mu}^2+\hat{p}_{\nu}^2\right)\right]
  -\hat{p}_{\mu}^3\hat{p}_{\nu}
  -\hat{p}_{\mu}\hat{p}_{\nu}^3\right\}
  \noenum
  \nexteq{2.0ex}
  {\phantom{\Delta_{\mu\nu}(p)={}}}
  +\lambda_0(\hat{p}_{\mu}-c_1\hat{p}_{\mu}^3)
  (\hat{p}_{\nu}-c_1\hat{p}_{\nu}^3),
  \enum
  \nexteq{3.0ex}
  \hat{p}^2=\sum_{\mu}\hat{p}_{\mu}^2,
  \quad
  \hat{p}^4=\sum_{\mu}\hat{p}_{\mu}^4,
  \enum
}
is then obtained. The associated Faddeev--Popov action
is diagonal in momentum space too, the inverse propagator
being $\Delta(p)=\hat{p}^2-c_1\hat{p}^4$ in this case.

\subsection 3.3 Gauge-field propagator

The normalization condition (2.6) and the requirement that
the coefficient $c_0$ must be positive imply the
strict positivity of the matrix $\Delta_{\mu\nu}(p)$
for all momenta $p$ summed over in eq.~(3.14).
In particular, the gauge-fixed action has
no zero-modes and the gauge-field propagator
in momentum space is given by
\equation{
  \langle \tilde{A}^a_{\mu}(p)\tilde{A}^b_{\nu}(q)^{\ast}\rangle=
  \frac{1}{2}TL^3\delta^{ab}\delta_{pq} D_{\mu\nu}(p)+\rmO(g_0),
  \enum
}
where $D_{\mu\nu}(p)$ denotes the (matrix) inverse of $\Delta_{\mu\nu}(p)$.
Recalling the Fourier representation (3.2),(3.3) of the gauge potential,
it is straightforward to pass from here to the propagator in
position space.

\section 4. The gradient-flow coupling

Apart from some minor technical details,
the definition of the gradient-flow coupling in the
theory with \openSF boundary conditions is the same as
in the case of periodic
and SF boundary conditions considered in
refs.~[\ref{NogradiEtAl}] and [\ref{FritzschRamos}],
respectively.
All formulae in this section refer to the lattice
theory, but their form in the continuum limit should be
quite obvious (see
refs.~[\ref{WilsonFlow},\ref{Mainz}] for an introduction to
the subject).

\subsection 4.1 Flow equation

The Yang--Mills gradient flow
evolves the gauge field as a function of
a parameter $t\geq0$, which is referred to as the flow time.
On the lattice,
the link variables $V_t(x,\mu)$ at flow time $t$ are determined by
the boundary condition
\equation{
  \left.V_t(x,\mu)\right|_{t=0}=U(x,\mu)
  \enum
}
and the first-order evolution equation
\equation{
  \partial_tV_t(x,\mu)V_t(x,\mu)^{-1}=-w_{x,\mu}g_0^2
  (\partial^a_{x,\mu}\Sw)(V_t)T^a,
  \enum
}
where $\Sw$ denotes the Wilson action, i.e.~the
gauge action (2.5) with $c_0=\cG=\cG'=1$ and $c_1=0$.
The gradient of the action on the right of the flow equation
is defined in appendix A and the
weight factor $w_{x,\mu}$ is given by
\equation{
  w_{x,\mu}=\cases{2 & if $x_0=0$ and $\mu>0$,\cr
                   \noalign{\vskip1ex}
                   0 & if $x_0=T$ and $\mu>0$,\cr
                   \noalign{\vskip1ex}
                   1 & otherwise.\cr}
  \enum
}
Assigning weight 2 to the spatial links at $x_0=0$
ensures that the gradient flow does not generate O($a$) lattice effects
at positive flow time
[\ref{TMopenQCD}]. At $x_0=T$, on the other hand,
the weight factor is set to zero so that
the boundary values of the gauge field are preserved.

\subsection 4.2 Solution of the flow equation to leading order in $g_0$

In perturbation theory, the link variables
\equation{
  V_t(x,\mu)=\exp\{g_0B_{\mu}(t,x)\},
  \qquad
  \left.B_{\mu}(t,x)\right|_{t=0}=A_{\mu}(x),
  \enum
}
are parameterized by a $t$-dependent
gauge potential $B_{\mu}(t,x)$. Furthermore,
a gauge-damping term must be added on the right of
the flow equation (4.2), as otherwise the gauge modes
of the field are not guaranteed to remain small
in the course of the flow-time evolution.

A possible choice of the gauge-damping term is
\equation{
  \alpha_0\{V_t(x,\mu)\omega_t(x+\hat{\mu})V_t(x,\mu)^{-1}-\omega_t(x)\},
  \enum
  \nexteq{2.5ex}
  \omega_t(x)=g_0\sum_{\nu}\drvstar{\nu}B_{\nu}(t,x),
  \enum
}
where $\alpha_0$ (the ``gauge-damping parameter'') can be set
to any positive value\kern1pt\footnote{$\dagger$}{\footnotefont%
At the boundaries of the lattice,
the derivatives $\sdrvstar{0}B_0(t,x)$ in eq.~(4.6) assume the
gauge potential $B_0(t,x)$ to be defined at all integer times $x_0$
through its Fourier representation. This amounts to setting
$\sdrvstar{0}B_0(t,x)=2B_0(t,x)$ at $x_0=0$ and
$\sdrvstar{0}B_0(t,x)=0$ at $x_0=T$.}.
As in the continuum theory [\ref{WilsonFlow}],
the structure of the gauge-damping term is such that it
can be removed by a $t$-dependent gauge transformation.
Gauge-invariant combinations
of the link variables at fixed flow time are therefore
not affected by the inclusion of the term in the flow equation.

At tree-level of perturbation theory, the flow equation
is linear and can be solved analytically by passing to
momentum space. Explicitly, the Fourier components
of the gauge potential $B_{\mu}(t,x)$ are given by
\equation{
  \tilde{B}_{\mu}(t,p)=\sum_{\nu}\biggl\{
  {\rme^{-t\hat{p}^2}\over\hat{p}^2}
  (\hat{p}^2\delta_{\mu\nu}-\hat{p}_{\mu}\hat{p}_{\nu})\tilde{A}_{\nu}(p)
  +{\rme^{-\alpha_0t\hat{p}^2}\over\hat{p}^2}
  \hat{p}_{\mu}\hat{p}_{\nu}\tilde{A}_{\nu}(p)\biggr\}+\rmO(g_0)
  \enum
}
to leading order in the gauge coupling.

\subsection 4.3 Running coupling

Let $G_{\mu\nu}(t,x)$ be the standard clover lattice expression for
the gauge-field tensor at flow time $t$
(see ref.~[\ref{WilsonFlow}], for example).
The expectation value of its square,
\equation{
  E(t,x)=\frac{1}{4}\sum_{\mu,\nu}G^a_{\mu\nu}(t,x)G^a_{\mu\nu}(t,x),
  \enum
}
does not require renormalization and
has a regular perturbation expansion in powers of $g_0^2$ without
constant term [\ref{WilsonFlow},\ref{RenFlow}].
In infinite volume and at vanishing quark masses,
the dimensionless combination
\equation{
  \gbarinf^2=\kfactinf
  \left\{t^2\langle E(t,x)\rangle\right\}_{\sqrt{8t}=r},
  \enum
}
with the normalization constant $\kfactinf$ chosen such that
\equation{
  \gbarinf^2=g_0^2+\rmO(g_0^4),
  \enum
}
has therefore all the properties of a renormalized coupling that
runs with the smoothing range $r$ of the gradient flow.

Renormalized couplings can be defined in exactly the same way in finite volume,
but in order to obtain a coupling that runs with $L$,
the ratios $T/L$, $x_0/L$ and $r/L$ must be set to some
fixed values (as before the quark masses are assumed to vanish).
In this paper the coupling is taken to be
\equation{
  \gbar^2=k\left\{t^2\langle E(t,x)\rangle
  \right\}_{T=L,x_0=L/2,\sqrt{8t}=cL},
  \enum
}
where $c$ is a dimensionless parameter that is left unspecified
for the moment.
Some numerical studies of the theory with SF
boundary conditions suggest that the
statistical variance and lattice-spacing-dependence of the coupling
are both reasonably small on the accessible lattices
if $c$ is in the range $0.3-0.5$ [\ref{FritzschRamos}].
The normalization constant $k$ is again determined by requiring
$\gbar^2$ to coincide with $g_0^2$ in the weak-coupling limit
(see appendix B).

\section 5. Topology-freezing and autocorrelation times

On lattices with periodic boundary conditions,
the space of gauge fields is connected, but
effectively divides into sectors of fixed topological
charge close to the continuum limit, with
the field space ``between the sectors'' being suppressed
in the functional integral by a large power of the lattice
spacing [\ref{WilsonFlow},\ref{BrunoSommer}]. Simulations of QCD tend to get
trapped in the emerging sectors and may consequently
lead to biased results.

The issue has been extensively studied in the past
(see refs.~[\ref{DelDebbioTauQ},\ref{SchaeferTauQ}], in particular),
and the situation is not fundamentally
different if SF boundary conditions are chosen.
The topological sectors exist in this case too [\ref{SF}]
and so does the topology-freezing problem [\ref{FixedQ}].

\subsection 5.1 Definition of the topological charge $Q$

Since the topological sectors are strictly separated from
each other only in the continuum limit, the definition of
the topological charge on the lattice requires some choices
to be made. As shown in ref.~[\ref{WilsonFlow}],
the emergence of the sectors can be quantitatively
understood by transforming the QCD functional integral
to an integral over the gauge field at gradient-flow
time $t>0$. A straightforward discretization
of the topological density then leads to the definition
\equation{
  Q={1\over32\pi^2}
  \sum_x\sum_{\mu,\nu}
  \dual{G}^a_{\mu\nu}(t,x)G^a_{\mu\nu}(t,x),
  \enum
}
which is automatically consistent with the
division of the field space in the continuum limit if
$t$ is set to some fixed value in physical units.
Moreover, the moments $\langle Q^n\rangle$ of the charge
distribution do not
require renormalization and can be shown to be independent of $t$
up to lattice effects of order $a^2$ [\ref{WilsonFlow}].
In this section, $t$ is scaled with $L$ such that
$\sqrt{8t}=0.3\times L$, but other reasonable choices of $t$
would make no difference in the following.

When SF or \openSF boundary conditions are imposed,
the definition (5.1) needs to be complemented
with a prescription of how exactly the charge
density is defined at time $0$ and $T$.
In the continuum theory, the density vanishes
at these times
and the boundary points may therefore just as well be
excluded from the sum over $x$ in eq.~(5.1)
(for reasons given later,
further time-slices near the open boundary will be omitted
when studying the autocorrelations of $Q$).

\subsection 5.2 Topology-freezing with SF boundary conditions

For illustration, it may be worth considering a sample
simulation of the Schr\"odinger functional, where the
topological charge of the gauge field is practically
unchanged for very long periods of simulation time
(see fig.~3). In this run, the O($a$)-improved
theory [\ref{SW},\ref{SFimp}] with gauge action (2.5)
and three flavours of nearly massless quarks was simulated%
\kern1pt\footnote{$\dagger$}{\footnotefont%
The parameters of the gauge action were
$\beta=3.81$, $c_0=5/3$ and $c_1=-1/12$.
For the coefficient of the Sheikholeslami--Wohlert term in
the quark action, the non-perturbatively determined value
$\csw=1.635109$ [\ref{CswThree}] was used.
The coefficients of the O($a$) boundary counterterms
were all set to their tree-level values.
At the chosen value of the quark hopping parameter,
$\kappa=0.137119$, the current-quark masses are
practically equal to zero on this lattice
[\ref{CAThree}].}.
All simulations reported in this paper
were performed with a version of the HMC algorithm [\ref{HMC}],
as described in ref.~[\ref{TMopenQCD}] and
implemented in the
publicly available {\tt openQCD} program package [\ref{OQCD}].

\topinsert
\vbox{
\vskip0.0cm
\centerline{\epsfxsize=10.5 true cm\epsfbox{plots/Qhist.eps}}
\vskip0.3cm
\figurecaption{%
History of the topological charge in three-flavour QCD
on a $36\times24^3$ lattice with SF (black line)
and \openSF (grey line) boundary conditions,
plotted as a function of the simulation time in units
of molecular-dynamics time (see subsect.~5.2 for further
details).
}
}
\endinsert

At the chosen point in parameter space,
the spatial size $L$ of the lattice is about $1.2$ fm [\ref{CswThree}].
The lattice considered is thus representative of the situation
typically encountered at the low-energy end of a step-scaling iteration,
where at most one or two further steps need to be taken before
contact with the large-volume regime of the theory can safely be made.
As is evident from fig.~3, the topological charge sectors
are poorly sampled in this case.
Moreover, when the
lattice spacing is reduced at fixed $L$, the associated
autocorrelation times grow very rapidly and the
simulation algorithm will, in practice, be trapped
in a single charge sector.

The loss of ergodicity of the simulations compromises
their correctness and must be expected to affect the calculated quantities
in various ways.
Provided the simulation algorithm gets trapped in the charge zero sector,
the issue may be less relevant on lattices with sizes $T,L\ll1$ fm,
where the QCD functional integral is totally
dominated by the gauge fields in a neighborhood of the classical
vacuum configuration. In the continuum theory and at fixed $T/L$,
the variance $\langle Q^2\rangle$
of the topological charge is expected to be a
universal function of the gradient-flow coupling, which
decreases roughly like
\equation{
  \langle Q^2\rangle\propto\gbar^{2\nu}\rme^{-b/\gbar^2}
  \enum
}
with $L$.
The simulation data for the variance shown in
fig.~4 actually reach fairly small values
already at $L\simeq1$ fm, but for
the charge fluctuations to be completely negligible, the
box size may have to be as small as $0.3$ fm or even
smaller. In practice, step-scaling iterations thus necessarily
pass through at least a few values of $L$, where
the contributions of the non-trivial charge sectors
and the topology-freezing issue cannot be ignored.

\topinsert
\vbox{
\vskip0.0cm
\centerline{\epsfxsize=6.6 true cm\epsfbox{plots/Qsq.eps}}
\vskip0.3cm
\figurecaption{%
Variance of the topological charge $Q$ (data points) in the
$\SUthree$ gauge theory on a $24^4$ lattice with SF boundary conditions,
plotted as a function of the inverse of
the gradient-flow coupling (4.11) with $c$
set to $0.3$.
The full line is a fit to the data of the form (5.2),
while the dotted line indicates where the data would have
to lie if the susceptibility $\langle Q^2\rangle/L^4$
were independent of the volume.
}
}
\endinsert

\subsection 5.3 Autocorrelation times from large to small volumes

With \openSF boundary conditions, the topological charge is
not quantized and the field space remains connected in the continuum
limit. There is therefore no reason to expect the simulations
to become effectively non-ergodic when the lattice spacing is
taken to zero. Whether the change from SF to \openSF
boundary conditions is profitable in practice however depends
on whether the relevant autocorrelation times are significantly reduced
or not.

In order to ensure a fair comparison,
the summation range in the definition (5.1) of the topological charge
on the lattices with \openSF boundary conditions
is chosen so as to maximize the integrated autocorrelation
times of $Q$ and thus excludes the time-slices near the
open boundary, where the charge density fluctuates more rapidly than
elsewhere in the space-time volume [\ref{openQCD},\ref{GlynnMawhinney}].
The integrated autocorrelation time of $Q$ then turns out to be
larger than the one of all other observables considered and is
probably close to the leading exponential autocorrelation time.

\topinsert
\ninepoint
\newdimen\digitwidth
\setbox0=\hbox{\rm 0}
\digitwidth=\wd0
\catcode`@=\active
\def@{\kern\digitwidth}
\tenpoint
\tablecaption{Integrated autocorrelation times in molecular-dynamics units
on a $24^4$ lattice}
\vskip-3.0ex

$$\vbox{
\settabs\+&xxxxxxxi&
          &xxxxxxxxxxxxxxxxxxxxxxxxxxxxxi&xxxi&
          &xxxxxxxxxxxxxxxxxxxxxxxxxxx&\cr
\thicktablerule
\vskip1.2ex
        \+&&&\hfill SF\hfill&&&\hfill \openSF\hfill&\cr
\vskip-1.3ex
        \+&&&\leaders\hrule&&&\leaders\hrule&\cr
\settabs\+&xxxxxx&x&
           xxxxxx&&
           xxxxxxixx&i&
           xxxxxxix&x&
           xxxxxxi&xxx&
           xxxxxxi&&
           xxxxxxx&i&
           xxxxxxx&x&
           xxxxxx&\cr
\vskip0.2ex
   \+& \hfill $\beta$\hfill
    && \hfill $\gbar^2$\hfill
    && \hfill $\tauint(Q)$\hfill
    && \hfill $\tauint(Q^2)$\hfill
    && \hfill $\tauint(\Ebar)$\hfill
    && \hfill $\gbar^2$\hfill
    && \hfill $\tauint(Q)$\hfill
    && \hfill $\tauint(Q^2)$\hfill
    && \hfill $\tauint(\Ebar)$\hfill
    &\cr
\vskip1.0ex
\thintablerule
\vskip1.2ex\ninepoint
   \+& \hfill $5.96$\hfill
    && \hfill $44.1(2)$\hfill
    && \hfill $49(6)$\hfill
    && \hfill $25(2)$\hfill
    && \hfill $25(2)$\hfill
    && \hfill $39.9(2)$\hfill
    && \hfill $51(7)$\hfill
    && \hfill $30(4)$\hfill
    && \hfill $24(2)@$\hfill
    &\cr
\vskip0.3ex
   \+& \hfill $6.09$\hfill
    && \hfill $26.5(1)$\hfill
    && \hfill $154(22)$\hfill
    && \hfill $70(8)$\hfill
    && \hfill $46(5)$\hfill
    && \hfill $24.2(1)$\hfill
    && \hfill $152(29)$\hfill
    && \hfill $@79(12)$\hfill
    && \hfill $45(6)@$\hfill
    &\cr
\vskip0.3ex
   \+& \hfill $6.21$\hfill
    && \hfill $17.8(1)$\hfill
    && \hfill $386(51)$\hfill
    && \hfill $189(24)$\hfill
    && \hfill $74(8)$\hfill
    && \hfill $16.5(1)$\hfill
    && \hfill $297(47)$\hfill
    && \hfill $138(12)$\hfill
    && \hfill $62(6)@$\hfill
    &\cr
\vskip0.3ex
   \+& \hfill $6.42$\hfill
    && \hfill $@9.0(1)$\hfill
    && \hfill $1595(199)$\hfill
    && \hfill $1138(162)$\hfill
    && \hfill $315(40)$\hfill
    && \hfill $9.63(2)$\hfill
    && \hfill $438(66)$\hfill
    && \hfill $199(19)$\hfill
    && \hfill $93(7)@$\hfill
    &\cr
\vskip0.3ex
   \+& \hfill $6.59$\hfill
    && \hfill $$\hfill
    && \hfill $$\hfill
    && \hfill $$\hfill
    && \hfill $$\hfill
    && \hfill $6.92(3)$\hfill
    && \hfill $176(23)$\hfill
    && \hfill $125(14)$\hfill
    && \hfill $74(8)@$\hfill
    &\cr
\vskip0.3ex
   \+& \hfill $7.00$\hfill
    && \hfill $$\hfill
    && \hfill $$\hfill
    && \hfill $$\hfill
    && \hfill $$\hfill
    && \hfill $4.21(2)$\hfill
    && \hfill $58(7)$\hfill
    && \hfill $35(4)$\hfill
    && \hfill $31(3)@$\hfill
    &\cr
\vskip0.3ex
   \+& \hfill $7.50$\hfill
    && \hfill $$\hfill
    && \hfill $$\hfill
    && \hfill $$\hfill
    && \hfill $$\hfill
    && \hfill $2.97(1)$\hfill
    && \hfill $31(2)$\hfill
    && \hfill $23(1)$\hfill
    && \hfill $26(2)@$\hfill
    &\cr
\vskip0.3ex
   \+& \hfill $30.00@$\hfill
    && \hfill $$\hfill
    && \hfill $$\hfill
    && \hfill $$\hfill
    && \hfill $$\hfill
    && \hfill $0.24(1)$\hfill
    && \hfill $@6(1)$\hfill
    && \hfill $@4(1)$\hfill
    && \hfill $@3(1)@$\hfill
    &\cr
\vskip1.2ex
\thicktablerule
}
$$
\vskip0.0ex
\endinsert

In the case of the three-flavour theory considered at the beginning
of subsect.~5.2, the beneficial effect of choosing \openSF boundary
conditions is evident (see fig.~3). Reliable determinations
of autocorrelation times however require much longer runs
than reported in fig.~3.
The results of some extensive simulations of the $\SUthree$ gauge
theory with Wilson plaquette action are listed in table~1.
While the lattice size was held fixed in these calculations,
the inverse gauge coupling $\beta=6/g_0^2$ and thus the box
size $L$ in physical units varies over a range of values.
Apart from the integrated autocorrelation times of $Q$, $Q^2$ and
\equation{
  \Ebar=L^{-3}\sum_{\mib x}\left.E(t,x)\right|_{x_0=L/2},
  \enum
}
the gradient-flow coupling is quoted for both
SF and \openSF boundary conditions. All observables were
evaluated at flow time $t$ corresponding to $\sqrt{8t}/L=0.3$.

In the range of $\beta$ from $5.96$ to $6.42$, the box size $L$
decreases from about $2.4$ to $1.2$ fm.
All autocorrelation times listed in table~1
grow rapidly in this range, particularly so in the
case of SF boundary conditions. At smaller values of
$L$, simulations of the Schr\"odinger functional
tend to be trapped in a sector of fixed topological charge
and the determination of the autocorrelation times becomes
impractical.
With \openSF boundary conditions, on the other hand,
the simulations do not show any sign of a freezing effect.
Moreover, towards the perturbative regime,
the calculated autocorrelation times decrease and reach
fairly small values at $\gbar^2\leq 3$.

\subsection 5.4 Scaling behaviour and statistical errors

The autocorrelation times quoted in table~1
were measured on a fixed lattice. If the
gradient-flow coupling is held fixed instead,
the autocorrelation times must be expected to grow
with the lattice size.
They tend to increase
very rapidly in the case of SF boundary conditions
as soon as the topology-freezing effects set in
[\ref{DelDebbioTauQ},\ref{SchaeferTauQ}].
With \openSF boundary conditions, the autocorrelation times
probably grow approximately like $1/L^2$ [\ref{openQCD}],
except perhaps at very small coupling,
where the theory is nearly Gaussian
and the autocorrelation times may conceivably
grow more slowly [\ref{RenHMC}].

Since the variance of the gradient-flow coupling is a renormalized quantity
of order $\gbar^4$, the number of {\it statistically
independent}\/ measurements of the coupling required for a given
relative statistical error is expected to be practically
independent of the lattice spacing and only slowly varying with
the physical size of the lattice. The empirical studies reported
in this section confirm this and moreover show that, with only 100
measurements, the values of the coupling quoted in table~1
could be reproduced to a precision of about $1\%$ ($0.6\%$ deep
in the perturbative regime).

\section 6. Boundary lattice effects

Boundary lattice effects decrease only linearly
with the lattice spacing and are therefore potentially large.
By tuning the coefficients
of the appropriate boundary counterterms,
they can in principle be reduced to O($a^2$),
but so far this has only been possible
at low orders of perturbation theory.

As explained below, the expectation values of
observables localized in the center of the lattice
tend to be weakly affected by the boundary lattice effects.
In practice, and if only such observables are considered,
a fine-tuning of the improvement coefficients is then
unnecessary.

\subsection 6.1 How large are the boundary lattice effects?

A first impression of how important the boundary lattice
effects are can be obtained by studying the dependence
of the quantities of interest on the
boundary improvement coefficients.
The data plotted in fig.~5, for example,
show that the gradient-flow coupling depends only
weakly on $\cG$ and $\cG'$ at the points in parameter space
considered. On these lattices, and
if the values of the coefficients
where O($a$)-improvement is achieved are in the range $[0.5,1.5]$,
the boundary lattice effects at $\cG=\cG'=1$ are thus
at most one percent or so.

\topinsert
\vbox{
\vskip0.0cm
\centerline{\epsfxsize=11.5 true cm\epsfbox{plots/Sens.eps}}
\vskip0.1cm
\figurecaption{%
Sensitivity of the gradient-flow coupling $\gbar^2$
to changes of the improvement
coefficients $\cG$ (left column) and $\cG'$ (right column)
in the $\SUthree$ gauge theory
with Wilson action and \openSF boundary conditions.
All data points were obtained on $24^4$ lattices at $\beta=7.5$
and are normalized so that the fitted
straight lines pass through zero at $\cG=\cG'=1$.
In the upper and lower rows of plots, the coefficient $c$ [eq.~(4.11)]
was set to $0.3$ and $0.2$, respectively. The associated
central values of the coupling are $2.97$ and $2.53$.
}
\vskip0.1cm
}
\endinsert

Deep in the perturbative regime, the improvement coefficients are
close to their values at tree-level of perturbation theory.
In the O($a$)-improved theory with $\SUthree$ gauge group,
Wilson gauge action and $\Nf$ quark flavours, for example,
the one-loop correction in
\equation{
  \cG'=1+\left\{-0.08900(5)+0.0191410(1)\times\Nf\right\}g_0^2
  +\rmO(g_0^4)
  \enum
}
is indeed reasonably small [\ref{OldGbar},\ref{SintSommer}].
The gradient-flow coupling (at, say, $c\leq0.3$ and $L\geq16$)
is therefore expected to be affected
by O($a$) boundary effects at a level of at most a small fraction of a
percent in this regime
if the improvement coefficients are set to their
tree-level values.
Clearly, the situation may be less favourable at large couplings
and non-perturbative scaling studies are required to be able
to exclude the presence of significant residual O($a$) effects
(see subsect.~6.3).

\subsection 6.2 Finite-size effects and the large-volume limit

The simulation results shown in fig.~5
can be theoretically understood, to some extent at least,
by studying the volume dependence of the expectation values
of observables $\obs(x)$ localized in the center of the space-time
volume.
In the present context, the observable of interest
is $E(t,x)$ at some flow-time $t$, but the argumentation
in the following paragraphs applies to any observable $\obs(x)$
with vacuum quantum numbers and fixed localization range.

Finite-volume effects are universal and are best discussed
directly in the continuum theory. Clearly,
differentiation of $\langle\obs(x)\rangle$
with respect to $T$ amounts to inserting the Euclidean expression
for the Hamilton operator in the expectation value.
At fixed $T/L$, the differential relation
\equation{
  {\partial\langle\obs(x)\rangle\over\partial L}=
  \frac{1}{2}\sum_{\mib y}
  \left\langle\obs(x)
  \left\{T_{00}(y)|_{y_0=0}+T_{00}(y)|_{y_0=T}\right\}
  \right\rangle_{\rm c}+\ldots
  \enum
}
is thus obtained, where $T_{\mu\nu}(y)$ denotes the energy-momentum tensor
and the ellipsis stands for three further terms in which one
of the space coordinates plays the r\^ole of time. As indicated by
the subscript ``c'', the two-point functions in this formula are
the connected parts of the full correlation functions.

When probed by local fields at distances
significantly larger than its localization range, as is the case in eq.~(6.2)
if $L$ is large, the observable $\obs(x)$ behaves like a
strictly local field of dimension $d\geq4$
[\ref{RenFlow},\ref{Mainz}].
Moreover, in the perturbative regime of QCD,
the dilation symmetry is only broken by logarithms of the scale factor
(as before, the quark masses are assumed to vanish).
The $L$-dependent parts of $\langle\obs(x)\rangle$
are therefore expected to fall off approximately like $L^{-d}$
at large $L$ until the large-volume regime of the theory is reached,
where scale invariance is strongly violated.
At this point, the remaining finite-volume effects
however tend to be already much smaller
than $\langle\obs(x)\rangle$ to the extent that
they can usually be neglected%
\kern1pt\footnote{$\dagger$}{\footnotefont%
In the theory with two or more flavours of massless quarks, the
asymptotic form of the finite-volume effects can be worked out
in chiral perturbation theory [\ref{BaerGolterman}].
The leading terms are of order $L^{-6}$ if periodic boundary conditions
are chosen and otherwise of order $L^{-4}$.
}.

\topinsert
\vbox{
\vskip0.0cm
\centerline{\epsfxsize=10.0 true cm\epsfbox{plots/ELoo.eps}}
\vskip0.3cm
\figurecaption{%
Volume dependence of $k_{\infty}t^2\langle E(t,x)\rangle$
at $x_0=L/2$ and fixed $t$.
The black data points
were obtained by simulating the $\SUthree$ gauge theory
at $\beta=7.5$ on a range of $L^4$ lattices
with \openSF boundary conditions and tree-level improved
Wilson gauge action.
The chosen value of the flow time $t$ is such that
$\sqrt{8t}/L=0.3$ and $\gbar^2=2.501(3)$ at $L=16$.
}
}
\endinsert

Numerical simulations confirm that
the expectation value of $E(t,x)$
converges to its infinite-volume limit in this way
(see fig.~6).
In the range $L\geq4\sqrt{8t}$, and
within statistical errors of about one per mille,
the data plotted in fig.~6
lie on a straight line that extrapolates to
$\gbarinf^2=2.538(2)$
(grey line and data point). Moreover,
even on the smallest lattice considered, the expectation value
differs from its infinite-volume limit by less than $2\%$.

\subsection 6.3 Synthesis

The volume dependence
of the expectation values $\langle\obs(x)\rangle$
and their dependence on
the improvement coefficients $\cG$ and $\cG'$
are closely related to each other.
Differentiation of $\langle\obs(x)\rangle$
with respect to the latter actually
leads to expressions like the one in eq.~(6.2).
The derivatives are therefore expected to be of order
$1/L^5$ in the perturbative regime.
In particular,
the contribution of the O($a$) boundary counterterms
to the gradient-flow coupling
is proportional to $t^2/L^5$, which explains the difference
(by about a factor $5$)
of the data in the upper and lower row of plots in fig.~5.

\topinsert
\vbox{
\vskip0.0cm
\centerline{\epsfxsize=9.0 true cm\epsfbox{plots/Sigma.eps}}
\vskip0.3cm
\figurecaption{%
Dependence on the lattice spacing of the
step-scaling function $\Sigma(2,u,1/L)$ in the
tree-level O($a$)-improved $\SUthree$ gauge theory
at $u=2.968$ (black data points). In this calculation,
the Wilson gauge action was used and the coefficient $c$ [eq.~(4.11)]
was set to $0.3$.
A linear extrapolation of the plotted data
yields the value $\sigma(2,u)=4.49(2)$
for the step-scaling function
in the continuum theory (grey data point).
}
}
\endinsert

Eventually, the importance of the O($a$) boundary effects
should be assessed by performing suitable scaling tests.
The step-scaling function $\Sigma(s,u,1/L)$
of the gradient-flow coupling, for example,
is expected to approach its continuum limit
with a rate proportional to $1/L^2$ (rather than $1/L$)
if there are no significant O($a$) effects
[\ref{StepScaling},\ref{OldGbar}].
Consistently with the theoretical discussion
in this section and the reported empirical results,
the data shown in fig.~7 are in fact compatible with
the residual O($a$) boundary effects being smaller than
the statistical errors.

In practice the choice of the parameter $c=\sqrt{8t}/L$ of
the gradient-flow coupling may require some tuning.
At small values of $c$,
the O($a$) boundary effects are strongly suppressed,
but the smoothing range $\sqrt{8t}$ must also
be significantly larger than the lattice spacing in order
to avoid large O($a^2$) contributions to the expectation value of
$E(t,x)$.
Increasing $T/L$ may be worth considering
if these criteria cannot both be met on the accessible lattices.

\section 7. Concluding remarks

The combination of observables and  boundary conditions
advertised in this paper provides a technically
attractive framework for step-scaling studies.
Since the to\-po\-logy-freezing issue is avoided,
non-perturbative computations of Symanzik improvement coefficients
(such as those recently reported in [\ref{CswThree},\ref{CAThree}])
are likely to profit from the use
of \openSF boundary conditions as well.
The fact that the gauge and ghost
propagators assume a simple form
moreover suggests that numerical stochastic
perturbation theory is straightforward to implement
with these boundary conditions
(see ref.~[\ref{DiRenzoScorzato}] for a review
of numerical stochastic perturbation theory and
[\ref{BrambillaBridaEtAl},\ref{BridaHesse}] for
the latest advances in this field).

In QCD many observables suitable for step scaling
can be constructed using the extension of the
Yang--Mills gradient flow to the quark fields [\ref{ChFlow}].
Unlike the case of periodic boundary conditions [\ref{MonahanOrginos}],
the chosen observables should preferably be localized in
the central region of the space-time volume as otherwise
the O($a$)-improvement of the calculated quantities
may be complicated by boundary lattice effects.
The normalization of the renormalized
quark field $\ren{\chi}(t,x)$ at flow-time $t>0$, for example,
may be fixed by requiring the expectation value
of the observable
\equation{
  \obs(x)=
  \ren{\chibar}(t,x)(\slash{D}-\lvec{\slash{D}})\ren{\chi}(t,x)
  \enum
}
at $x_0=L/2$
to be equal to its value at tree-level of perturbation
theory [\ref{EMflowIII}]. Local fields at vanishing flow time
can then be renormalized by probing them with suitable products
of renormalized fields at $t>0$.

The fact that
the gradient-flow coupling $\gbarinf^2$ in infinite volume
can be accurately computed at all flow times visited
in the course of a step-scaling calculation is intriguing.
Extrapolations to
the continuum and the infinite-volume limit are required in
this computation, but both limits are reached
fairly rapidly at fixed flow time and could actually be
taken simultaneously.
Current correlation functions in position space
and other quantities of interest might be accessible
in this way too, provided they depend on
a single external scale that can play the r\^ole of the flow time.

\vskip0.4ex
I am indebted to John Bulava for helpful correspondence
on three-flavour QCD with SF boundary conditions
and the topology-freezing issue in this theory.
Thanks also go to Agostino Patella and Stefan Schaefer
for interesting discussions on
various topics addressed in this paper.
All simulations were performed on a
dedicated PC cluster at CERN. I am grateful
to the CERN management for funding this machine
and to the CERN IT Department for technical support.

\appendix A. Notational conventions

The gauge group is taken to be $\SUn$ with $N=3$ in
sections 5 and 6.
In any basis $T^a$, $a=1,\ldots,N^2-1$, of
complex, anti-Hermitian and traceless $N\times N$ matrices,
the general element of the Lie algebra of $\SUn$ is given by
$X^aT^a$ with real components $X^a$
(repeated group indices are automatically summed over).
The generators are assumed to satisfy
\equation{
  \tr\{T^aT^b\}=-\frac{1}{2}\delta^{ab}
  \enum
}
but are otherwise left unspecified.
Lorentz indices $\mu,\nu,\ldots$
range from $0$ to $3$ and are {\it not}\/ automatically
summed over when they occur in matching pairs.

The differential operators $\partial^a_{x,\mu}$ act on
differentiable functions $f(U)$ of the gauge field $U$ according to
\equation{
  \partial^a_{x,\mu}f(U)=
  {\rmd\over\rmd s}f(\rme^{sX}U)\!\left.{{\vphantom{r\over s}}}\right|_{s=0},
  \hskip1.7em
  X(y,\nu)=\cases{T^a & if $(y,\nu)=(x,\mu)$,\cr
                          \noalign{\vskip0.8ex}
                        0  & otherwise.\cr}
  \enum
}
While these operators depend on the choice of the generators $T^a$,
the gradient field $T^a\partial^a_{x,\mu}f(U)$
can be shown to be basis-independent.

\appendix B. Normalization of the gradient-flow coupling

The computation of the normalization factor
\equation{
  k^{-1}=\bigl\{t^2{\cal E}_0(t,x)\bigr\}_{T=L,x_0=L/2,\sqrt{8t}=cL}
  \enum
}
in eq.~(4.11) requires the expectation value
\equation{
  \langle E(t,x)\rangle={\cal E}_0(t,x)g_0^2+\rmO(g_0^4)
  \enum
}
to be calculated at tree-level of perturbation theory.
In the lattice theory with the Wilson plaquette action and
SF boundary conditions,
the calculation was recently performed by
Fritzsch and Ramos [\ref{FritzschRamos}].
The results quoted below hold for all
actions considered in this paper and cover the case of
both \openSF and SF boundary conditions.

\subsection B.1 Open-SF boundary conditions

To leading order in the gauge coupling,
$E(t,x)$ is a quadratic expression in the gauge
potential at flow time $t$ and thus, via eq.~(4.7), of the
fundamental field.
Recalling eq.~(3.17), the result
\equation{
  {\cal E}_0(t,x)=
  \noenum
  \nexteq{2.0ex}
  \qquad
  {N^2-1\over TL^3}\sum_p
  \rme^{-2t\hat{p}^2}
  \biggl\{\sin(p_0x_0)^2\sum_{l=1}^3S_{l0}(p)
  +\cos(p_0x_0)^2\sum_{l>j=1}^3S_{lj}(p)\biggr\}
 \enum
}
is then obtained, where
\equation{
  S_{\mu\nu}(p)=(1-\frac{1}{4}\hat{p}_{\mu}^2)(1-\frac{1}{4}\hat{p}_{\nu}^2)
  \left\{\hat{p}_{\mu}^2D_{\nu\nu}(p)+\hat{p}_{\nu}^2D_{\mu\mu}(p)
  -2\hat{p}_{\mu}\hat{p}_{\nu}D_{\mu\nu}(p)\right\}.
  \enum
}
Substitution of these equations in eq.~(B.1) yields the
normalization factor in the form of a momentum sum that can
be evaluated numerically.

\subsection B.2 SF boundary conditions

For these boundary conditions, the normalization factor
is again given by eq.~(B.1), but the expression for
\equation{
  {\cal E}_0(t,x)=
  \noenum
  \nexteq{2.0ex}
  \qquad
  {N^2-1\over TL^3}{\sum_p}'
  \rme^{-2t\hat{p}^2}
  \biggl\{\cos(p_0x_0)^2\sum_{l=1}^3S_{l0}(p)
  +\sin(p_0x_0)^2\sum_{l>j=1}^3S_{lj}(p)\biggr\}
 \enum
}
differs from the one obtained in the case of \openSF boundary conditions.
In particular, the sines and cosines
in the curly bracket are interchanged
with respect to eq.~(B.3) and
the time component of the momentum,
\equation{
  p_0=n_0{\pi\over T}, \qquad n_0=0,1,\ldots,T-1,
  \enum
}
is summed over a different set of values.
The primed summation symbol in eq.~(B.5) indicates that the terms at $p_0=0$
are given the weight $1/2$ and that the $p=0$ term is to be omitted.

\subsection B.3 Explicit form of the gauge-field propagator

Equation (B.4) involves the gauge-field propagator $D_{\mu\nu}(p)$
and is therefore not fully explicit. Starting from the
inverse propagator (3.15), a few lines of algebra however show that
\equation{
  D_{\mu\nu}(p)=
  {\delta_{\mu\nu}\over r_{\mu}}
  -{v_{\mu}v_{\nu}\over1+\sum_{\rho}v_{\rho}^2r_{\rho}}+
  (\lambda_0^{-1}-1)
  {\hat{p}_{\mu}\hat{p}_{\nu}\over(\hat{p}^2-c_1\hat{p}^4)^2},
  \enum
}
where
\equation{
  r_{\mu}=\hat{p}^2-c_1(\hat{p}^4+\hat{p}^2\hat{p}_{\mu}^2),
  \qquad
  v_{\mu}=c_1\hat{p}_{\mu}^3/r_{\mu}.
  \enum
}
Note that the second term in eq.~(B.7) is of order
$a^4$ in the lattice spacing, while the last term is a gauge term
that cancels in eq.~(B.4).

\beginbibliography


\bibitem{StepScaling}
M. L\"uscher, P. Weisz, U. Wolff,
{\it A numerical method to compute the running coupling in asymptotically
free theories},
Nucl. Phys. B359 (1991) 221


\bibitem{WilsonFlow}
M. L\"uscher,
{\it Properties and uses of the Wilson flow in lattice QCD},
JHEP 08 (2010) 071 [Erratum: {\it ibid.} 03 (2014) 092]


\bibitem{RenFlow}
M. L\"uscher, P. Weisz,
{\it Perturbative analysis of the gradient flow in non-Abelian gauge
theories},
JHEP 02 (2011) 051


\bibitem{ChFlow}
M. L\"uscher,
{\it Chiral symmetry and the Yang--Mills gradient flow},
JHEP 04 (2013) 123


\bibitem{NogradiEtAl}
Z. Fodor, K. Holland, J. Kuti, D. Nogradi, C. H. Wong,
{\it The Yang--Mills gradient flow in finite volume},
JHEP 11 (2012) 007;
{\it The gradient flow running coupling scheme},
PoS (Lattice 2012) 050

\bibitem{FritzschRamos}
P. Fritzsch, A. Ramos,
{\it The gradient flow coupling in the Schr\"odinger functional},
JHEP 10 (2013) 008;
{\it Studying the gradient flow coupling in the Schr\"odinger functional},
PoS (Lattice 2013) 319

\bibitem{RamosTwisted}
A. Ramos,
{\it The gradient flow in a twisted box},
PoS (Lattice 2013) 053

\bibitem{JarnoMWT}
J. Rantaharju,
{\it The gradient-flow coupling in minimal walking technicolor},
PoS (Lattice 2013) 084


\bibitem{EMflowI}
H. Suzuki,
{\it Energy-momentum tensor from the Yang--Mills gradient flow},
PTEP 2013 (2013) 083B03

\bibitem{EMflowIb}
L. Del Debbio, A. Patella, A. Rago,
{\it Space-time symmetries and the Yang--Mills gradient flow},
JHEP 11 (2013) 212

\bibitem{EMflowII}
M. Asakawa et al. (FlowQCD Collab.),
{\it Thermodynamics of SU(3) gauge theory from gradient flow},
arXiv:1312.7492

\bibitem{EMflowIII}
H. Makino, H. Suzuki,
{\it Lattice energy-momentum tensor from the Yang--Mills gradient flow
--- inclusion of fermion fields},
PTEP 2014 (2014) 063B02


\bibitem{Mainz}
M. L\"uscher,
{\it Future applications of the Yang--Mills gradient flow in lattice QCD},
PoS (Lattice 2013) 016


\bibitem{SF}
M. L\"uscher, R. Narayanan, P. Weisz, U. Wolff,
{\it The Schr\"odinger functional ---
a renormalizable probe for non-Abelian gauge theories},
Nucl. Phys. B384 (1992) 168

\bibitem{SFquark}
S. Sint,
{\it On the Schr\"odinger functional in QCD},
Nucl. Phys. B421 (1994) 135


\bibitem{FixedQ}
P. Fritzsch, A. Ramos, F. Stollenwerk,
{\it Critical slowing down and the gradient flow coupling in the
Schr\"odinger functional},
PoS (Lattice 2013) 461


\bibitem{openQCD}
M. L\"uscher, S. Schaefer,
{\it Lattice QCD without topology barriers},
JHEP 07 (2011) 036


\bibitem{Wilson}
K. G. Wilson, {\it Confinement of quarks}, Phys. Rev. D10 (1974) 2445


\bibitem{Iwasaki}
Y. Iwasaki,
{\it Renormalization group analysis of lattice theories and
improved lattice action. II -- Four-dimensional non-Abelian SU(N) gauge model},
preprint UTHEP-118 (1983) and arXiv:1111.7054


\bibitem{OnShell}
M. L\"uscher, P. Weisz,
{\it On-shell improved lattice gauge theories},
Commun. Math. Phys. 97 (1985) 59 [E: {\it ibid.} 98 (1985) 433]


\bibitem{SFaction}
S. Aoki, R. Frezzotti, P. Weisz,
{\it
Computation of the improvement coefficient $c_{\rm sw}$ to 1-loop
with improved gluon actions},
Nucl. Phys. B540 (1999) 501


\bibitem{TMopenQCD}
M. L\"uscher, S. Schaefer,
{\it
Lattice QCD with open boundary conditions and twisted-mass reweighting},
Comput. Phys. Commun. 184 (2013) 519


\bibitem{BrunoSommer}
M. Bruno, R. Sommer,
{\it On the $N_f$-dependence of gluonic observables},
PoS (Lattice 2013) 321


\bibitem{DelDebbioTauQ}
L. Del Debbio, H. Panagopoulos, E. Vicari,
{\it $\theta$-dependence of SU(N) gauge theories},
JHEP 08 (2002) 044

\bibitem{SchaeferTauQ}
S. Schaefer, R. Sommer, F. Virotta,
{\it Investigating the critical slowing down of QCD simulations},
PoS(LAT2009)032;
{\it Critical slowing down and error analysis in lattice QCD simulations},
Nucl. Phys. B845 (2011) 93


\bibitem{SW}
B. Sheikholeslami, R. Wohlert,
{\it Improved continuum limit lattice action for QCD with Wilson fermions},
Nucl. Phys. B259 (1985) 572

\bibitem{SFimp}
M. L\"uscher, S. Sint, R. Sommer, P. Weisz,
{\it Chiral symmetry and O(a) improvement in lattice QCD},
Nucl. Phys. B478 (1996) 365


\bibitem{CswThree}
J. Bulava, S. Schaefer,
{\it Improvement of $N_f=3$ lattice QCD with Wilson\hfill\break
fermions and tree-level improved gauge action},
Nucl. Phys. B874 (2013) 188


\bibitem{CAThree}
J. Bulava, M. Della Morte, J. Heitger, C. Wittemeier,
{\it Determinantion of $\ca$ in three-flavour lattice QCD with
Wilson fermions and tree-level improved gauge action},
PoS (Lattice 2013) 311


\bibitem{HMC}
S. Duane, A. D. Kennedy, B. J. Pendleton, D. Roweth,
{\it Hybrid Monte Carlo},
Phys. Lett. B195 (1987) 216


\bibitem{OQCD}
{\tt http://cern.ch/luscher/openQCD}


\bibitem{GlynnMawhinney}
G. McGlynn, R. D. Mawhinney,
{\it Scaling, topological tunneling and actions
for weak coupling DWF calculations},
PoS (Lattice 2013) 027


\bibitem{RenHMC}
M. L\"uscher, S. Schaefer,
{\it Non-renormalizability of the HMC algorithm},
JHEP 04 (2011) 104


\bibitem{OldGbar}
M. L\"uscher, R. Sommer, P. Weisz, U. Wolff,
{\it A precise determination of the running coupling in the SU(3)
Yang--Mills theory},
Nucl. Phys. B413 (1994) 481

\bibitem{SintSommer}
S. Sint, R. Sommer,
{\it The running coupling from the QCD Schr\"odinger functional:
a one-loop analysis},
Nucl. Phys. B465 (1996) 71


\bibitem{BaerGolterman}
O. B\"ar, M. Golterman,
{\it Chiral perturbation theory for gradient-flow observables},
Phys. Rev. D89 (2014) 034505


\bibitem{DiRenzoScorzato}
F. Di Renzo, L. Scorzato,
{\it Numerical stochastic perturbation theory for full QCD},
JHEP 10 (2004) 073

\bibitem{BrambillaBridaEtAl}
M. Brambilla, M. Dalla Brida, F. Di Renzo, D. Hesse, S. Sint,
{\it Numerical stochastic perturbation theory in the Schr\"odinger functional},
PoS (Lattice 2013) 325

\bibitem{BridaHesse}
M. Dalla Brida, D. Hesse,
{\it Numerical stochastic perturbation theory and the gradient flow},
PoS (Lattice 2013) 326


\bibitem{MonahanOrginos}
C. Monahan, K. Orginos,
{\it Finite volume renormalization scheme for fermionic operators},
PoS (Lattice 2013) 443

\endbibliography

\bye